\documentclass{article}
\usepackage[english]{babel}
\usepackage{amsmath,amssymb,theorem}

\newcommand{\nobracket}{}

\newcommand{\tmop}[1]{\ensuremath{\operatorname{#1}}}
{\theorembodyfont{\rmfamily}}

\begin{document}

\title{ Generalized Mandelstam-Leibbrandt regularization }
\author{J. Alfaro \\
	Facultad de F\'\i sica, Pontificia Universidad Cat\'olica de Chile,\\
	Casilla 306, Santiago 22, Chile.\\
	jalfaro@uc.cl}

\maketitle

\begin{abstract}
	Algebraic non-covariant gauges are used often in string theory, Chern-Simons theory, gravitation  and gauge theories. Loop integrals, however, have spurious singularities that need to be regularized. The most popular and consistent regularization is the Mandelstam- Leibbrandt(ML) prescription. 
  This paper  extends the ML prescription outside the light cone. It shares all the properties of 
  light-cone ML regularization: It preserves naive power counting and gauge
  invariance. Moreover, using dimensional regularization(DR), we get a closed
  form for the basic integrals, including divergent and finite pieces. These results simplify calculations in gauge theories and open new avenues for applications in non-local models.
\end{abstract}

\section{Introduction}

The modern description of particles and interactions in the subatomic world has been based on gauge theories. In a gauge theory we have redundant degrees of freedom that need to be removed by a gauge choice, which is a constraint imposed upon the gauge fields. The election of a gauge can simplify significantly the computation of Green functions.

For several years, various groups of researchers have been working on the
quantization of gauge theories in algebraic non-covariant
gauges{ {\cite{Lei},\cite{soldati}}}. These gauges became popular, because
they simplify the analysis of gauge theories due to: a) the decoupling  of
Fadeev-Popov ghost contributions{\footnote{There are some
		subtleties related to  this point. Please see {\cite{Lei}} chapter 4.4.}} b) starting at the classical level only
physical degrees of freedom are present and c) absence of Gribov ambiguities.
Non-covariant gauges have been used to discuss Yang-Mills (YM) theories \cite{YM},
super-symmetric (YM) theories \cite{man}, super-gravity and super-strings\cite{gs}.

The main difficulty of algebraic non-covariant gauges is that Feynman
integrals have singularities in the $k_0$ complex plane, $k_{\mu}$ being the
loop variable being integrated over. Several regularization have been tested
to deal with this situation. But they present problems and inconsistencies.

One possible way to regularize the singularity is  use of the principal
value prescription(PV). But in this case, naive power counting is lost and
double poles in DR are present already at the one loop level, introducing
logarithmic dependence on the external momenta in the divergent part of the
integral. The lost of naive power counting is related to the fact that using
PV regularization, Wick's rotation is not allowed.

$\alpha$-prescription\cite{alpha} is a deformation of the gluon propagator in the temporal gauge to regularize the integrals.
But, it does not preserve the Ward identities of the gauge theory\cite{Lei}.

Mandelstam-Leibbrandt(ML) prescription \cite{man,lei}, was introduced to deal with these problems
in the light-cone gauge (lcg): It preserves naive power counting, gauge
invariance and  Wick's rotation from Minkowski to
Euclidean space is justified. 

Also, renormalization in the lcg can be implemented using ML\cite{soldati}.

Moving away from the lcg presents problems though. Although a generalization of ML
in lcg was implemented for other axial-type gauges, a canonical derivation of
the prescription is lacking. Besides the classification of counter-terms and
the renormalization in these gauges is problematic{\cite{soldati}}.

To understand some  problems of the Standard Model
(SM) such as the origin of neutrino masses and  oscillations, Very
Special Relativity(VSR) has been proposed\cite{vsr}. In this approach, the 4 parameters
$\tmop{Sim} (2)$ subgroup of the Lorentz group is assumed to be the symmetry
of nature. $\tmop{Sim} (2)$  is characterized by changing a fixed null vector
$n_{\mu}$ by a scale factor, so ratios $\frac{n \cdot p_1}{n \cdot p_2}$ where
$p_i$ are particle momenta, are $\tmop{Sim} (2)$ invariant but break Lorentz
invariance. Such non-local terms permits the introduction of chiral neutrino
masses \cite{vsr2} and gauge invariant masses for photon \cite{phmass}and graviton\cite{alsan}.

In VSR, the propagators and vertexes of fermions and bosons are allowed by $\tmop{Sim} (2)$
to have the aforementioned non-local terms. In loop computations  these
non-local terms share the same singularities of algebraic non-covariant gauges. 

Few years ago, we derived complete formulas for ML regularized integrals in
the lcg{\cite{alfaro}}. 
A scale symmetry and a regularity condition are enough to determine the
integrals in closed form, using DR. These formulas have been fundamental to define a gauge, $Sim(2)$ invariant regularization for VSR models, with interesting phenomenological predictions \cite{alfaro2}.

In this paper we implement the same program for algebraic non-covariant
gauges, outside the light-cone. We derived the integrals by leveraging a scale symmetry and imposing a regularity condition. They preserve naive
power counting and the divergences are polynomials in the external momenta.
They preserve gauge invariance. Moreover, they have a smooth light cone limit.

It turns out that the integrals agree with the ones computed using the
generalized ML prescription.

In section 2, we review the computation of integrals in the lcg. In section 3,
we compute the single spurious pole integrals. Section 4 contains the
computation of integrals with higher order spurious poles. It remains to solve
a recurrence relation. This is done in section 5. In section 6, we draw some conclusions.
Appendix A lists some basic
loop integrals. In Appendix B, we mention the existence of a symmetry of
single spurious poles integrals. In Appendix C, a computer routine to solve
the recurrence relation introduced in section 4 is presented. Appendix D deals with the calculation of finite integrals in two-dimensional space-time, that 
serve as a test of our results. In Appendix E we compute some integrals to compare with \cite{Lei},\cite{soldati}.

\section{ML prescription in the light-cone gauge.}

We are assuming a Minkowski metric $\eta_{\mu \nu} = \tmop{diag} (1, - 1, - 1,
- 1)$.

The light cone gauge is
\begin{eqnarray*}
  n^{\mu} A_{\mu} = 0, & n\cdot n = 0 & 
\end{eqnarray*}
$A_\mu$	is a gauge field.	

In loop calculations spurious singularities appear. A typical loop integral such as\footnote{Here as well as the rest of the paper  $m^2$ actually means $m^2-i\epsilon$ with $\epsilon>0$.}:
\[ \int d p \frac{1}{[p^2 + 2 p\cdot q - m^2]^a} \frac{1}{(n\cdot p)^b} \]
has a singularity when $n\cdot p = 0$.

These singularities have been treated using various prescriptions: Principal value(PV), $\alpha$-prescription\cite{Lei}.
All of these prescriptions have some problems: $\alpha$- prescription does not preserve the Ward identities of the gauge symmetry. PV does not permit the rotation to Euclidean space. As a consequence naive power counting is lost.

To illustrate the problem of PV prescription, we recall the integral defined in equation (6.1) of \cite{soldati}.
\begin{equation}
	I = \int \frac{d p}{p^2 (p - k)^2 p^+}, p^+ = \frac{p^0 + p^3}{\sqrt{2}}\label{1}
\end{equation}
PV is defined by:
\begin{equation}
	\frac{1}{p^+} = \lim_{\varepsilon \to 0} \frac{1}{2} \left(
	\frac{1}{p^+ + i \varepsilon} + \frac{1}{p^+ - i \varepsilon} \right)
\end{equation}
Naive power counting would imply that $I$ is finite. Instead, the result for the divergent part of $I$, using PV is:
\begin{equation}
	\mathbb{P}I_{\tmop{PV}} = i (- \pi)^{\omega} \frac{2 \omega - 3}{\omega - 2}
	\frac{(k^2)^{\omega - 2}}{k^+} \frac{\Gamma (2 - \omega) [\Gamma (\omega -
		1)]^2}{\Gamma (2 \omega - 2)}
\end{equation}
Here $\omega = \frac{d}{2}$, where $d$ is the space-time dimension in DR.

We see that $\mathbb{P}I_{\tmop{PV}}$ contains a double pole at $\omega = 2$,
which is unusual for a one loop integral. Worse yet, the single pole of
$\mathbb{P}I_{\tmop{PV}}$ exhibits a log dependence on the external momentum
$k_{\mu}$. This will create problems in the renormalization of the theory.

The source of the problem mentioned above is that PV puts the poles in the second and third
quadrants of the $p_0$ complex plane(for $p^3 > 0$). So the Wick's rotation to
Euclidean space pick up an extra factor, from the residue of the encircled
pole.

To solve the inconsistencies of  previous prescriptions, the ML regularization introduces a second null vector $\bar{n}_{\mu}$ and
define:
 \begin{equation}
	\frac{1}{n \cdot p} = \lim_{\varepsilon \rightarrow 0} \frac{\bar{n}
		\cdot p}{n \cdot p \bar{n} \cdot p + i \varepsilon},\bar{n}\cdot\bar{n}=0 \label{ml}
\end{equation} 

In a given Lorentz frame, we can choose $n_\mu=(n_0,\vec{n}),\bar{n}_\mu=(n_0,-\vec{n})$ with $n_0^2-\vec{n}\cdot\vec{n}=0$. The poles in the $p_0$ complex plane are situated following the same pattern as  covariant poles (the poles are in the II and IV quadrants) such that the Wick's rotation from Minkowski to Euclidean space does not find any encircled pole. It follows that ML preserves naive power counting of  loop integrals\footnote{In Appendix E we compute integral (\ref{1}) using ML. It is finite.}
. Moreover,  in gauge
theories, it maintains the Ward identities of the gauge symmetry.

Without loss of generality, we can impose the condition $\bar{n}\cdot n=1$.

A disadvantage of ML is that calculations are very long compared with usual DR in covariant gauges. Moreover, explicit formulas for the ML integrals, including finite and divergent parts,  were not readily available.

In {\cite{alfaro}} we developed a technique to calculate such integrals based
on a symmetry and a regularity condition. We got a closed form for the
integrals, using DR. This is a great simplification compared with previous
methods that computed mainly the pole part because the  calculation of the finite part was
complicated.

To illustrate  the techniques of \cite{alfaro}, 
let us compute the following simple integral:
\begin{equation}
	A_{\mu} = \int d p \frac{f (p^2) p_{\mu} }{n\cdot p}
\end{equation}
where $f$ is an arbitrary function.$d p$ is the integration measure in $d$
dimensional space and $n_{\mu}$ is a fixed null vector($n\cdot n = 0$). The integrand is  singular when $n\cdot p = 0$.

To compute $A_{\mu}$, using ML, we have to know the specific form of $f$, provide a
 form of  $n_{\mu}$, $\bar{n}_{\mu}$, and evaluate the residues of
all poles of \ $\frac{f (p^2) }{n.p}$ in the $p_0$ complex plane, a rather formidable task for an arbitrary $f$.

Instead, we notice the following symmetry:
\begin{equation}
	n_{\mu} \rightarrow \lambda n_{\mu}, \bar{n}_{\mu} \rightarrow \lambda^{- 1}
	\bar{n}_{\mu}, \lambda \neq 0, \lambda \varepsilon R \label{symmetry}
\end{equation}
The properties of $n_{\mu}$ and $\bar{n}_{\mu}$ are preserved:
\begin{eqnarray*}
	0 = n \cdot n \rightarrow \lambda^2 n \cdot n = 0 &  & \\
	0 = \bar{n} \cdot \bar{n} \rightarrow \lambda^{- 2} \bar{n} . \cdot \bar{n}
	= 0 &  & \\
	1 = n \cdot \bar{n} \rightarrow n \cdot \bar{n} = 1 &  & 
\end{eqnarray*}
We see from (\ref{ml}) that:
\begin{equation}
	\frac{1}{n\cdot p} \rightarrow \frac{1}{n\cdot p} \lambda^{- 1}
\end{equation}
Now we compute $A_{\mu}$, using its symmetries. It is a Lorentz vector
which scales under (\ref{symmetry}) as $\lambda^{- 1}$. The only Lorentz
vectors that are available in this case are $n_{\mu}$ and $\bar{n}_{\mu}$. But
(\ref{symmetry}) forbids $n_{\mu}$. That is:
\[ A_{\mu} = a \bar{n}_{\mu} \]

Multiply by  $n^{\mu}$ to find $A \cdot n = a$. Thus $a = \int d p f (p^2)$.
Finally, we get:
\begin{equation}
	\int d p \frac{f (p^2) p_{\mu} }{n \cdot p} = \bar{n}_{\mu} \int d p f
	(p^2)
\end{equation}
We consider now a more general integral. Regularity of
the answer will determine it uniquely.

\begin{equation}
	A = \int d p \frac{H (p^2, p \cdot q)}{n \cdot p} = \bar{n} \cdot q f (q^2,
	n \cdot q \bar{n} \cdot q) \label{generic}
\end{equation}
$q_{\mu}$ is an external momentum, a Lorentz vector. $H$ is an arbitrary
function. The last relation follows from (\ref{symmetry}), for a certain $f$
we will find next.

Derive respect to $q^\mu$ and multiply by $n^\mu$
\begin{eqnarray}
	\frac{\partial A}{\partial q^{\mu}} n^{\mu} = \int d p H_{, u} = & g (x) = &
	\nonumber\\
	f (x, y) + 2 y \frac{\partial}{\partial x} f (x, y) + y
	\frac{\partial}{\partial y} f (x, y) &  &  \label{pdi1}
\end{eqnarray}
We defined $u = p \cdot q$, $x = q^2$, $y = n \cdot q \bar{n} \cdot q$. $()_{,u}$ means derivative respect to $u$.

Assuming that the solution and its partial derivatives are finite in the neighborhood of $y = 0$, it follows from the equation that $f (x, 0) = g (x)$.
That is the partial differential equation has a unique regular solution.
 In the next chapter, we will explain how to solve this type of equations, using the method of characteristics \cite{copson}. In the same way we solved 
 equation (\ref{pdi1}) in \cite{alfaro}.

 	Now we apply equation (\ref{pdi1}) to compute integrals that appear in gauge theory loops:
 	\begin{equation}
 		\int dp \frac{1}{[p^2 + 2 p \cdot q - m^2]^a}  \frac{1}{n \cdot p} =
 		\bar{n} \cdot q f (x, y)
 	\end{equation}
 	In this case we have:
 	\begin{equation}
 		 g (x) = - 2 a \int dp \frac{1}{[p^2 - x - m^2]^{a + 1}}
 	\end{equation}
 	The unique regular solution of (\ref{pdi1}) is:
 	\begin{eqnarray*}
 		f (x, y) = - \frac{1}{y}  \left\{ \int dp [p^2 - x - m^2]^{- a} - \int dp
 		[p^2 - x + 2 y - m^2]^{- a} \right\} &  & 
 	\end{eqnarray*}
 	We can check  that $f (x, 0) = - 2 a \int dp [p^2 - x - m^2]^{- a - 1} = g
 	(x)$.
 	The remaining integral can be computed using DR. The result is:
 	\begin{eqnarray}
 		\int d p \frac{1}{[p^2 + 2 p\cdot q - m^2]^a} \frac{1}{n\cdot p} = (- 1)^{a + 1} i
 		(\pi)^{\omega} (- 2) \frac{\Gamma (a + 1 - \omega)}{\Gamma (a)}
 		\bar{n} \cdot q &  &  \nonumber\\
 		\int_0^{^1} d t \frac{1}{(m^2 + q\cdot q - 2 n\cdot q \bar{n}\cdot q t)^{a + 1 -
 				\omega}},  \omega = d / 2 &  \label{lcg1}
 	\end{eqnarray}
 	
 	The technique can be extended to obtain the general integral:
\begin{eqnarray}
  \int d p \frac{1}{[p^2 + 2 p\cdot q - m^2]^a} \frac{1}{(n\cdot p)^b} = (- 1)^{a + b} i
  (\pi)^{\omega} (- 2)^b \frac{\Gamma (a + b - \omega)}{\Gamma (a) \Gamma (b)}
  (\bar{n} \cdot q)^b &  &  \nonumber\\
  \int_0^{^1} d t t^{b - 1} \frac{1}{(m^2 + q\cdot q - 2 n\cdot q \bar{n}\cdot q t)^{a + b -
  \omega}},  \omega = d / 2 &  \label{lcg}
\end{eqnarray}
Here $d$ is the dimension of space-time in DR.

These results followed from the scale symmetry (\ref{symmetry}) plus the condition
that (\ref{lcg}) must be regular at $n\cdot q \bar{n}\cdot q = 0$.

In the next section we will apply the same method to the same type of
integrals for arbitrary $n\cdot n \neq 0, \bar{n} \cdot \bar{n} \neq 0$.

\section{Treatment of singularities outside the light cone. Single spurious pole.}

Numerous applications of algebraic non-covariant gauges to Yang-Mills theories, super-gravity and super-strings motivate the need to compute loop integrals with singularities of the type we described in the last section.

In this and the following section, we will calculate such integrals outside the lcg. We will find the whole answer, including divergent and finite parts.

In this section, integrals with a single spurious pole will be treated. They are somehow special in ML prescription and are the basis for  computation of integrals with higher order poles.

A central role will be played by a vector $F_\mu$ which is a function of $n_\mu,\bar{n}_\mu$. $F_\mu$ plays here the same role that $\bar{n}_\mu$ played in the lcg. Due to a consistency condition that we will derive below or a hidden symmetry that we present in Appendix B, $F_\mu$ must be a null vector, $F\cdot F=0$. This property determines the form of $F_\mu$.

The algebraic non-covariant gauge we are considering here is:
\begin{eqnarray*}
	n^{\mu} A_{\mu} = 0, & n\cdot n \neq 0 
\end{eqnarray*}
$A_\mu$	is a gauge field. This includes temporal gauge ($n\cdot n>0$)  and axial gauge ($n\cdot n<0$).

In loop calculations spurious singularities appear, which can be embedded in the following integral:
\begin{equation}
	A = \int d p \frac{H (p^2, p\cdot q)}{n\cdot p} 
\end{equation}

$d$ is the space-time dimension in DR.$H$ is an arbitrary function.

The generalized ML prescription is:
\begin{equation}
  \frac{1}{n\cdot p} = \lim_{\varepsilon \rightarrow 0} \frac{\bar{n}\cdot p}{n\cdot p \ 
  \bar{n}\cdot p + i \varepsilon}, \bar{n} \cdot \bar{n} \neq 0, n\cdot n \neq 0 \label{ML}
\end{equation}
In d=4, we can choose a Lorentz frame where $n_{\mu} = (n_0, 0, 0, n_3),
\bar{n}_{\mu} = (n_0, 0, 0, - n_3), n_0 \neq 0, n_3 \neq 0$.

Under the scale transformation:
\begin{eqnarray}
  n_{\mu} \rightarrow \lambda n_{\mu}, & \bar{n}_{\mu} \rightarrow \lambda^{-
  1} \bar{n}_{\mu}, & A \rightarrow \lambda^{- 1} A  \label{sca}
\end{eqnarray}
Let us assume the existence of a vector $F_{\mu} (n, \bar{n})$ that transforms
under (\ref{sca}) as
\begin{equation}
  F_{\mu} \rightarrow \lambda^{- 1} F_{\mu} \label{sF}
\end{equation}
We normalize it so that  $n\cdot F = 1$.

We will determine the exact form of $F_{\mu}$ later.

To simplify the notation, we define:
\begin{equation}
	x = \bar{n}	\cdot n, y = \bar{n} \cdot \bar{n}, 
	z = n\cdot n
\end{equation}

Introduce the following definitions:
\begin{equation}
	u = p\cdot q, v = q\cdot q, w = \frac{n\cdot q}{\sqrt{z}}, s = \sqrt{z} F\cdot q
\end{equation}
   This set provides a complete list of functions of $q_\mu$ which are Lorentz scalar and scale invariant  under (\ref{sca}).
   
Therefore, we can write in all generality:
\begin{equation}
	A = \int d p \frac{H (p^2, p\cdot q)}{n\cdot p} = F\cdot q f (v, w, s)
\end{equation}

That is:
\begin{eqnarray}
  \frac{\partial A}{\partial q^{\mu}} = \int d p \frac{H_{, u}
  p_{\mu}}{n\cdot p} =\nonumber\\
  F_{\mu} (f + s f_s) + 2 z^{- 1 / 2} s f_v q_{\mu} + z^{- 1} s f_w n_{\mu} &
  & 
\end{eqnarray}
$()_{,u}$ means derivative respect to $u$.

Multiplying by $n^\mu$, we get a partial differential equation for $f(v,w,s)$:
\begin{eqnarray}
  \frac{\partial A}{\partial q^{\mu}} n^{\mu} = \int d p H_{, u} = B (v) &  & 
  \nonumber\\
  (f + f_s s) + 2 s \omega f_v + s f_w = B (v) &  &  \label{f}
\end{eqnarray}
\[ f_s = \frac{\partial f}{\partial s}, f_v = \frac{\partial f}{\partial v},
   f_w = \frac{\partial f}{\partial w} \]
   
As in the lcg, we impose a regularity condition: $f_s, f_v, f_w$ must be finite at $s = 0$. From  equation (\ref{f}) we get $f (v,
w, 0) = B (v)$ . Thus, equation (\ref{f}) has a unique solution, which is
regular at $s = 0$.

To find the solution we use the method of characteristics{\cite{copson}}:
\begin{eqnarray}
  \begin{array}{lll}
    \dot{s} = s, & \dot{v} = 2 s w, & \dot{w} = s\\
    f + \dot{f} = B (v) &  & 
  \end{array} &  & 
\end{eqnarray}
\begin{eqnarray*}
  s = s_0 e^t, &  & \dot{w} = s_0 e^t, w = s_0 e^t + w_0,\\
  \dot{v} = 2 s_0^2 e^{2 t} + 2 s_0 w_0 e^t, &  & v = s_0^2 e^{2 t} + 2 s_0
  w_0 e^t + v_0\\
  w - s = w_0, &  & v + s^2 - 2 s w = v_0
\end{eqnarray*}
Homogeneous equation:
\begin{eqnarray*}
  \dot{f} = - f, & f = f_0 e^{- t} = f_0 s_0 \frac{1}{s} & 
\end{eqnarray*}

Thus, the general solution of the homogeneous($B = 0$) equation is:
\[ f_h = \frac{1}{s} \Pi (v + s^2 - 2 s w, w - s) \]

Where $\Pi$ is an arbitrary function.

A particular solution of the inhomogeneous equation is:
\begin{eqnarray*}
 f_p=a(t) e^{-t} ,\dot{a} e^{- t} = B (v (t)) &  & 
\end{eqnarray*}
\begin{equation}
  f_p = e^{- t} \int_{- \infty}^t d t' e^{t'} B (v (t')) \label{psol}
\end{equation}
Then, the most general solution of equation (\ref{f}) is:
\begin{equation}
	f = f_p + \frac{1}{s} \Pi (v + s^2 - 2 s w, w - s)
\end{equation}
Regularity at $s = 0$, implies $\Pi (v, w)$=0, all $v, w$.

Moreover, we see that:
\[ f_p (v, w, 0) = B (v) \]
Therefore $f_p$ defined in equation (\ref{psol}) is the unique regular solution to equation
(\ref{f}).

\subsection{Loop integrals}

Let us apply equation (\ref{psol}) to
\begin{equation}
L_1 = \int d p \frac{1}{[p^2 + 2 p\cdot q - m^2]^a} \frac{1}{n\cdot p}
\end{equation}	
We get:
\begin{equation}
	B (v) = - 2 a \int d p \frac{1}{[p^2 - v - m^2]^{a + 1}}
\end{equation}
Then, using equation (\ref{psol}), we obtain:
\begin{equation}
	f = - 2 a \int d p \int_0^1 d t \frac{1}{[p^2 - (v - 2 s w t + s^2 t^2) -
		m^2]^{a + 1}}
\end{equation}
Using dimensional regularization, we get:
\begin{eqnarray}
  \int d p \frac{1}{[p^2 + 2 p\cdot q - m^2]^a} \frac{1}{n\cdot p} = &  &  \nonumber\\
  2 (- 1)^a i \pi^{\omega} \frac{\Gamma \left( a + 1 - \omega
  \right)}{\Gamma (a)} F\cdot q \int_0^1 d t \frac{1}{(q\cdot q - 2 F\cdot q n\cdot q \ t + z
  (F\cdot q)^2 t^2 + m^2)^{a + 1 - \omega}} &  &  \label{onep}
\end{eqnarray}

We can see that naive power counting is preserved. Moreover, we recover the light cone integral (\ref{lcg}), when $n\cdot n =
\bar{n} \cdot  \bar{n} = 0$. To get this we require that $F_{\mu} \rightarrow
\frac{\bar{n}_{\mu}}{x}$ in the lcg limit. We should keep this condition in mind when we will derive the form of $F_\mu$ in the next subsection.

Other integrals can be obtained deriving respects to $q^{\mu}$. For instance:
\begin{eqnarray}
	\int d p \frac{1}{[p^2 + 2 p.q - m^2]^a} \frac{p_{\mu}}{n.p} = &  & 
	\nonumber\\
	(- 1)^a i \pi^{\omega} \frac{\Gamma \left( a - \frac{d}{2} \right)}{\Gamma
		(a)} F_{\mu} \int_0^1 d t \frac{1}{(q.q - 2 F.q n.q t + n.n (F.q)^2 t^2 +
		m^2)^{a - \frac{d}{2}}} &  &  \nonumber\\
	- 2 (- 1)^a i \pi^{\omega} \frac{\Gamma \left( a + 1 - \frac{d}{2}
		\right)}{\Gamma (a)} F.q \int_0^1 d t \frac{q_{\mu} - F_{\mu} n.q t - F.q
		n_{\mu} t + n.n F.q F_{\mu} t^2}{(q.q - 2 F.q n.q t + n.n (F.q)^2 t^2 +
		m^2)^{a + 1 - \frac{d}{2}}} &  & 
\end{eqnarray}

We integrate by part the second integral to obtain:
\begin{eqnarray}
	2 (- 1)^a i \pi^{\omega} \frac{\Gamma \left( a + 1 - \frac{d}{2}
		\right)}{\Gamma (a)} \int_0^1 d t \frac{F.q n.q t - n.n (F.q)^2 t^2}{(q.q -
		2 F.q n.q t + n.n (F.q)^2 t^2 + m^2)^{a + 1 - \frac{d}{2}}} = &  & 
	\nonumber\\
	2 (- 1)^a i \pi^{\omega} \frac{\Gamma \left( a - \frac{d}{2} \right)}{\Gamma
		(a)} \frac{1}{2} \int_0^1 d t t \frac{d}{d t} \frac{1}{(q.q - 2 F.q n.q t +
		n.n (F.q)^2 t^2 + m^2)^{a - \frac{d}{2}}} = &  &  \nonumber\\
	(- 1)^a i \pi^{\omega} \frac{\Gamma \left( a - \frac{d}{2} \right)}{\Gamma
		(a)} \frac{1}{(q.q - 2 F.q n.q + n.n (F.q)^2 + m^2)^{a - \frac{d}{2}}} - & 
	&  \nonumber\\
	(- 1)^a i \pi^{\omega} \frac{\Gamma \left( a - \frac{d}{2} \right)}{\Gamma
		(a)} \int_0^1 d t \frac{1}{(q.q - 2 F.q n.q t + n.n (F.q)^2 t^2 + m^2)^{a -
			\frac{d}{2}}} &  & 
\end{eqnarray}
So, finally, the integral is:
\begin{eqnarray}
	\int d p \frac{1}{[p^2 + 2 p \cdot q - m^2]^a} \frac{p_{\mu}}{n \cdot p} = &
	& \nonumber\\
	(- 1)^a i \pi^{\omega} \frac{\Gamma \left( a - \omega \right)}{\Gamma
		(a)} \frac{F_{\mu}}{(q \cdot q - 2 F \cdot q n \cdot q + z (F \cdot
		q)^2 + m^2)^{a - \omega}} &  & \nonumber\\
	- 2 (- 1)^a i \pi^{\omega} \frac{\Gamma \left( a + 1 - \omega
		\right)}{\Gamma (a)} F \cdot q \int_0^1 d t \frac{q_{\mu} - F \cdot q
		n_{\mu} t}{\left( q \cdot q - 2 F \cdot q \quad n \cdot q t + z (F \cdot
		q)^2 t^2 + m^2 \right)^{a + 1 - \omega}} &  & \label{pmu}
\end{eqnarray}
Integration by parts in a $t$ integral will be used several times in this work.

 But still we do not know what $F_\mu$ is. We will determine it in the next subsection.
\subsection{Consistency condition}

Consider the third derivative of equation (\ref{onep}) with respect to $q^{\mu}$, and evaluate it at $q^{\mu} = 0$. We get:
\begin{eqnarray}
	\int d p \frac{1}{[p^2 - m^2]^{a + 3}} \frac{p_{\mu} p_{\nu} p_{\lambda}}{n
		\cdot p} = \frac{1}{2} \frac{(- 1)^a i \pi^{\omega}}{\Gamma (a + 3)}
	\frac{\Gamma (a + 2 - \omega)}{(m^2)^{a + 2 - \omega}} &  &  \nonumber\\
	\{ \nobracket F_{\mu} \int_0^1 d t (\eta_{\nu \lambda} - t (n_{\lambda}
	F_{\nu} + F_{\lambda} n_{\nu}) + t^2 z F_{\lambda} F_{\nu}) + &  & 
	\nonumber\\
	F_{\nu} \int_0^1 d t (\eta_{\mu \lambda} - t (n_{\lambda} F_{\mu} +
	F_{\lambda} n_{\mu}) + t^2 z F_{\lambda} F_{\mu}) + &  &  \nonumber\\
	F_{\lambda} \int_0^1 d t (\eta_{\mu \nu} - t (n_{\nu} F_{\mu} + F_{\nu}
	n_{\mu}) + t^2 z F_{\nu} F_{\mu}) \} \nobracket &  & 
\end{eqnarray}

Let $d_{\tmop{ph}}$ be the physical dimension. Choose $a+2 =
\frac{d_{\tmop{ph}}}{2}$. Define $\varepsilon = d - d_{\tmop{ph}}$.Then the pole part ($\mathbb{P}$) is given by:

\begin{eqnarray}
	\mathbb{P} \int d p \frac{1}{[p^2 - m^2]^{\frac{d_{\tmop{ph}}}{2} + 1}}
	\frac{p_{\mu} p_{\nu} p_{\lambda}}{n\cdot p} = i (-
	1)^{\frac{^{d_{\tmop{ph}}}}{2}} \pi^{\frac{^{d_{\tmop{ph}}}}{2}}
	\frac{1}{\varepsilon} \frac{1}{\Gamma \left( \frac{d_{\tmop{ph}}}{2} + 1
		\right)} &  & \nonumber\\
	\{ F_{\mu} F_{\nu} F_{\lambda} z - F_{\mu} F_{\nu} n_{\lambda} - F_{\mu}
	F_{\lambda} n_{\nu} - F_{\lambda} F_{\nu} n_{\mu} + F_{\mu} \eta_{\nu
		\lambda} + F_{\nu} \eta_{\mu \lambda} + \nobracket F_{\lambda} \eta_{\mu
		\nu} \} \nobracket \label{2p3}
\end{eqnarray}

Contracting $\nu=\lambda$ we get:
\begin{eqnarray*}
  \mathbb{P} \int d p \frac{1}{[p^2 - m^2]^{\frac{d_{\tmop{ph}}}{2} + 1}}
  \frac{p_{\mu} p^2}{n\cdot p} = i (- 1)^{\frac{^{d_{\tmop{ph}}}}{2}}
  \pi^{\frac{^{d_{\tmop{ph}}}}{2}} \frac{1}{\varepsilon} \frac{1}{\Gamma
  \left( \frac{d_{\tmop{ph}}}{2} + 1 \right)} [F_{\mu} (d_{\tmop{ph}} + z
  F\cdot F) - F\cdot F n_{\mu}] &  & 
\end{eqnarray*}
Since naive power counting is preserved,  $\mathbb{P}$ can be computed
using (\ref{pmu}):
\begin{eqnarray*}
  \mathbb{P} \int d p \frac{1}{[p^2 - m^2]^{\frac{d_{\tmop{ph}}}{2} + 1}}
  \frac{p_{\mu} p^2}{n\cdot p} =\mathbb{P} \int d p \frac{1}{[p^2 -
  m^2]^{\frac{d_{\tmop{ph}}}{2}}} \frac{p_{\mu}}{n\cdot p} = F_{\mu} i (-
  1)^{\frac{^{d_{\tmop{ph}}}}{2}} \pi^{\frac{^{d_{\tmop{ph}}}}{2}}
  \frac{1}{\Gamma \left( \frac{d_{\tmop{ph}}}{2} \right)}
  \frac{2}{\varepsilon} &  & 
\end{eqnarray*}
To have consistency between the two ways of getting the pole part of the previous integral, we must have that:
\begin{eqnarray*}
  \frac{1}{d_{\tmop{ph}}} [F_{\mu} (d_{\tmop{ph}} + z F\cdot F) - F\cdot F n_{\mu}] =
  F_{\mu}, \tmop{or} &  & \\
  (F_{\mu} z - n_{\mu}) F\cdot F = 0 &  & 
\end{eqnarray*}
$F_{\mu} = \frac{n_{\mu}}{z}$ does not go to $\frac{\bar{n}_{\mu}}{\bar{n}
.n}$ when $z \rightarrow 0$. So the consistency condition implies $F\cdot F = 0$,
independently of the physical dimension $d_{\tmop{ph}}$.

To find $F_{\mu}$, we use Lorentz symmetry, plus the conditions $n\cdot F = 1$ and $F\cdot F= 0$.

We can write, for certain Lorentz scalars $a, b$:
\begin{eqnarray*}
	F_{\mu} = a \bar{n}_{\mu} + b \frac{n_{\mu}}{n \cdot n}, &  & \\
	n \cdot F = 1 \to & b = 1 - a \bar{n} \cdot n, &
	\\
	F \cdot F = 0 \to  a^2 \bar{n} \cdot \bar{n} + 2 a
	b \frac{\bar{n} \cdot n}{n \cdot n} + b^2 \frac{1}{n \cdot n} = 0, &  & \\
	a = \pm \frac{1}{\sqrt{(\bar{n} \cdot n)^2 - \bar{n} \cdot \bar{n} n \cdot
			n}}
\end{eqnarray*}
To fix the sign in $a$ we used the boundary condition $\lim_{z \rightarrow 0}
b = 0$, to recover the lcg result $F_{\mu} \rightarrow
\frac{\bar{n}_{\mu}}{x}$

That is:
\begin{eqnarray}
  F_{\mu} = \frac{\bar{n}_{\mu}}{D} - \frac{n_{\mu}}{z} \left( \frac{x}{D} - 1
  \right), & D = \frac{1}{\sqrt{x^2 - y z}} \label{Fmu}
\end{eqnarray}

$F_{\mu}$ scales as  equation (\ref{sF}) required.

We see that $F_{\mu}$ agrees with equation (A6.42) of {\cite{soldati}}.

As a further check that equation (\ref{pmu}) does provide the complete
integral, we compute a finite integral which is not listed in {\cite{Lei}},
{\cite{soldati}}. Choose $d_{\tmop{ph}} = 2$, $n_{\mu} = (n_0, n_1),
\bar{n}_{\mu} = (n_0, - n_1)$ and calculate:
\begin{equation}
	A_{\mu} = \int d p \frac{1}{[p^2 - m^2]^2} \frac{p_{\mu}}{n\cdot p}
\end{equation}
We get, using the generalized ML prescription (\ref{ML}):
\begin{eqnarray}
  \int d p \frac{1}{[p^2 - m^2]^2} \frac{p_1}{n\cdot p} = - i \pi \frac{n_1}{m^2}
  \left( \frac{1}{n_1 (n_0 + n_1)} \right) = i \pi \frac{1}{m^2} F_1 &  & \label{38}\\
  \int d p \frac{1}{[p^2 - m^2]^2} \frac{p_0}{n\cdot p} = \frac{i \pi n_0}{m^2}
  \frac{1}{n_0 (n_0 + n_1)} = i \pi \frac{1}{m^2} F_0 &  & \label{39}
\end{eqnarray}
which coincide with equation (\ref{pmu}) at $q_{\mu} = 0, a = 2$. Integrals (\ref{38},\ref{39}) are calculated in Appendix D.

It is easy to check that (\ref{onep}) includes all integrals
with a single power of $\frac{1}{n\cdot p}$ in Appendix 6 of {\cite{soldati}} or Appendix D of {\cite{Lei}}. In the aforementioned books the pole part of the integrals is listed, but equation (\ref{onep}) provides the whole integral for arbitrary values of $a,d$.

\section{Higher singularities,  $(\frac{1}{n\cdot p})^b$, $b\ge 2$.}

In the lcg the scaling symmetry and the regularity condition at $\bar{n}\cdot q n\cdot q=0$ were enough to determine the integral for $b\ge 2$. Outside the lcg, this is not so. 

But outside the lcg both $n_\mu$ and $\bar{n}_\mu$ are unconstrained, so it makes sense to compute derivatives with respects to $n_\mu$.

To proceed further we notice that:
\begin{eqnarray}
  \frac{\partial}{\partial n^{\nu}} \left( \frac{\bar{n} .p}{\bar{n} \cdot p n\cdot p +
  i \varepsilon} \right) = - \frac{(\bar{n} .p)^2 p_{\nu}}{(\bar{n} \cdot p n\cdot p + i
  \varepsilon)^2} = - \frac{p_{\nu}}{(n\cdot p)^2} &  &  \label{dnmu}
\end{eqnarray}

Using this identity in equation (\ref{onep}), we get:
\begin{eqnarray*}
  \int d p \frac{1}{[p^2 + 2 p\cdot q - m^2]^a} \frac{p_{\mu}}{(n\cdot p)^2} = &  & \\
  2 (- 1)^a i \pi^{\omega} \frac{\Gamma \left( a + 1 - \omega
  \right)}{\Gamma (a)} E_{\nu \mu} q^{\nu} \int_0^1 d t \frac{1}{(q\cdot q - 2 F\cdot q
  n\cdot q t + n\cdot n (F\cdot q)^2 t^2 + m^2)^{a + 1 - \omega}} + &  & \\
  2^2 (- 1)^a i \pi^{\omega} \frac{\Gamma \left( a + 2 - \omega
  \right)}{\Gamma (a)} F\cdot q &  & \\
  \int_0^1 d t \frac{E_{\nu \mu} q^{\nu} n\cdot q t - F\cdot q q_{\mu} t + n_{\mu}
  (F\cdot q)^2 t^2 - n\cdot n F\cdot q E_{\nu \mu} q^{\nu} t^2}{(q\cdot q - 2 F\cdot q n\cdot q t + n\cdot n
  (F\cdot q)^2 t^2 + m^2)^{a + 2 - \omega}} &  & 
\end{eqnarray*}

Where:
\begin{eqnarray}
  E_{\mu \nu} = \frac{x}{D^3} \bar{n}_{\mu} \bar{n}_{\nu} + \eta_{\mu \nu}
  \frac{y}{D (x + D)} + n_{\mu} n_{\nu} \frac{y^2}{D^2 (x + D)} \left(
  \frac{1}{D} + \frac{1}{x + D} \right) - (n_{\mu} \bar{n}_{\nu} + n_{\nu}
  \bar{n}_{\mu}) \frac{y}{D^3} &  &  \label{E}\\
  \frac{\partial F_{\mu}}{\partial n^{\nu}} = - E_{\mu \nu}, E_{\mu \nu} =
  E_{\nu \mu} &  &  \nonumber
\end{eqnarray}
$E_{\mu \nu}$ coincides with equation A6.44 of {\cite{soldati}}.

Consider \ the coefficient of $E_{\nu \mu} q^{\nu}$ in the second integral.
Integrating by parts in $t$, we get:
\begin{eqnarray}
  \int d p \frac{1}{[p^2 + 2 p\cdot q - m^2]^a} \frac{p_{\mu}}{(n\cdot p)^2} = &  & 
  \nonumber\\
  2 (- 1)^a i \pi^{\omega} \frac{\Gamma \left( a + 1 - \omega
  \right)}{\Gamma (a)}  \frac{E_{\nu \mu} q^{\nu}}{(q\cdot q - 2 F\cdot q n\cdot q + n\cdot n
  (F\cdot q)^2 + m^2)^{a + 1 - \omega}} + &  &  \nonumber\\
  2^2 (- 1)^a i \pi^{\omega} \frac{\Gamma \left( a + 2 - \omega
  \right)}{\Gamma (a)} \int_0^1 d t \frac{- (F\cdot q)^2 q_{\mu} t + n_{\mu}
  (F\cdot q)^3 t^2}{(q\cdot q - 2 F\cdot q n\cdot q t + n\cdot n (F\cdot q)^2 t^2 + m^2)^{a + 2 -
  \omega}} &  &  \label{2pmu}
\end{eqnarray}
Using equation (\ref{Fmu}) we get:
\begin{equation}
	E_{\mu \nu} =  F_{\mu} F_{\nu}(\rho z+1) + \rho (\eta_{\mu \nu} - n_{\mu}
	F_{\nu} - n_{\nu} F_{\mu}), \rho = \frac{y}{D (x + D)}\label{E2}
\end{equation}

And
\begin{equation}
	E_{\mu \nu} q^{\nu} = \frac{x}{D} (F\cdot q) F_{\mu} + \rho (q_{\mu} - n_{\mu}
	F\cdot q - n\cdot q_{} F_{\mu}) \label{Eq}
\end{equation}
Since the $b=2$ integral is well defined, we can write:
\begin{eqnarray}
  \frac{\partial}{\partial q^{\mu}} \int d p \frac{1}{[p^2 +
  2 p\cdot q - m^2]^{a - 1}} \frac{1}{(n\cdot p)^2} = - 2 (a - 1) \int d p \frac{1}{[p^2
  + 2 p\cdot q - m^2]^a} \frac{p_{\mu}}{(n\cdot p)^2} \label{45}
\end{eqnarray}

Equation (\ref{45}) defines the $b=2$ integral up to an additive constant, which is independent of $q_\mu$.

For a Lorentz invariant $g$, we can write:
\begin{eqnarray}
  \int d p \frac{1}{[p^2 + 2 p\cdot q - m^2]^{a - 1}}
  \frac{1}{(n\cdot p)^2} = g (v, w, s), &  & \nonumber\\
  \frac{\partial}{\partial q^{\mu}} g (v, w, s) = 2 g_v q_{\mu} + g_w
  \frac{n_{\mu}}{\sqrt{z}} + g_s F_{\mu} \sqrt{z} &  & \label{g0}
\end{eqnarray}
We use the notation:
\[ g_s = \frac{\partial g}{\partial s}, g_v = \frac{\partial g}{\partial v},g_w = \frac{\partial g}{\partial w} \]

From equations (\ref{2pmu},\ref{Eq},\ref{g0}) we get:
\begin{eqnarray}
  g_v = 2 \rho (- 1)^{a + 1} i \pi^{\omega} \frac{\Gamma \left( a + 1 -
  \omega \right)}{\Gamma (a - 1)}  \frac{1}{(q\cdot q - 2 F\cdot q n\cdot q + n\cdot n
  (F\cdot q)^2 + m^2)^{a + 1 - \omega}} & + &  \nonumber\\
  (2^2) i \pi^{\omega} (- 1)^a \frac{\Gamma \left( a + 2 - \omega
  \right)}{\Gamma (a - 1)} \int_0^1 d t \frac{(F\cdot q)^2 t}{(q\cdot q - 2 F\cdot q n\cdot q t +
  n\cdot n (F\cdot q)^2 t^2 + m^2)^{a + 2 - \omega}} &  &  \label{fv}
\end{eqnarray}

\begin{eqnarray}
	g_w = (- 1)^a 2^2 i \pi^{\omega}  \frac{\Gamma (a + 1 - \omega)}{\Gamma (a -
		1)}  \frac{\rho (F.q) \sqrt{z}}{(q \cdot q - 2 F \cdot qn \cdot q + n \cdot
		n (F \cdot q)^2 + m^2)^{a + 1 - \omega}} + &  &  \nonumber\\
	2^3  (- 1)^{a + 1} i \pi^{\omega}  \frac{\Gamma (a + 2 - \omega)}{\Gamma (a
		- 1)} \int_0^1 dt \frac{t^2 \sqrt{z}  (F \cdot q)^3}{(q \cdot q - 2 F \cdot
		qn \cdot qt + n \cdot n (F \cdot q)^2 t^2 + m^2)^{a + 2 - \omega}} &  & 
	\label{fw}
\end{eqnarray}
\begin{eqnarray}
	g_s = 2^2 (- 1)^{a + 1} i \pi^{\omega}  \frac{\Gamma (a + 1 -
		\omega)}{\Gamma (a - 1)} \frac{1}{\sqrt{z}}  \frac{\frac{x}{D}  (F \cdot q)
		- \rho n \cdot q }{(q \cdot q - 2 F \cdot qn \cdot q + n \cdot n
		(F \cdot q)^2 + m^2)^{a + 1 - \omega}} &  &  \label{fs}
\end{eqnarray}
Integrating $g_v$ over $v = q\cdot q$, we obtain :
\begin{eqnarray}
  g = \bar{g} (w, s) + 2 \rho (- 1)^a i \pi^{\omega} \frac{\Gamma \left( a -
  \omega \right)}{\Gamma (a - 1)}  \frac{1}{(q\cdot q - 2 F\cdot q n\cdot q + n\cdot n
  (F\cdot q)^2 + m^2)^{a - \omega}} + &  & \nonumber\\
  (2^2) (- 1)^{a + 1} i \pi^{\omega} \frac{\Gamma \left( a + 1 - \omega
  \right)}{\Gamma (a - 1)} \int_0^1 d t \frac{(F\cdot q)^2 t}{(q\cdot q - 2 F\cdot q n\cdot q t +
  n\cdot n (F\cdot q)^2 t^2 + m^2)^{a + 1 - \omega}} &  & \label{f2}
\end{eqnarray}

Since the integral exists there must be a $\bar{g} (w, s)$ that produce agreement between equations (\ref{f2},\ref{fw},\ref{fs}). Imposing the boundary condition that the integral vanishes when $v \to -\infty$,we get $\bar{g} (w, s) = 0$
\footnote{$g_v$ must be $\mathcal{O}(v^{- (1 + \delta)})$,$\delta>0$; and $g_s,g_w$ must vanish, at $v\to-\infty$.}.
As a further check, we used  Eq. (\ref{f2}) to obtain Eqs.(\ref{fw},\ref{fs}). We got that $\bar{g} (w, s)$ is a constant.

. We must have $a > \omega$. But the  integral is analytic almost everywhere, so the result can be extended by analytic continuation.

We have proved that:
\begin{eqnarray}
  \int d p \frac{1}{[p^2 + 2 p\cdot q - m^2]^a} \frac{1}{(n\cdot p)^2} = &  & 
  \nonumber\\
  2 \rho (- 1)^{a + 1} i \pi^{\omega} \frac{\Gamma (a + 1 - \omega)}{\Gamma
  (a)} \frac{1}{(q\cdot q - 2 F\cdot q n\cdot q + n\cdot n (F\cdot q)^2 + m^2)^{a + 1 - \omega}} +
  &  &  \nonumber\\
  (2^2) (- 1)^a i \pi^{\omega} \frac{\Gamma \left( a + 2 - \omega
  \right)}{\Gamma (a)} (F\cdot q)^2 \int_0^1 d t \frac{t}{(q\cdot q - 2 F\cdot q n\cdot q t + n\cdot n
  (F\cdot q)^2 t^2 + m^2)^{a + 2 - \omega}} &  &  \label{2p}
\end{eqnarray}
We see that in the light cone limit we recover equation(\ref{lcg}). Moreover,
we preserve the scaling property:
\begin{eqnarray*}
  \int d p \frac{1}{[p^2 + 2 p\cdot q - m^2]^a} \frac{1}{(n\cdot p)^2} = m^{d - 2 a - 2}
  \int d p \frac{1}{\left[ p^2 + 2 p. \frac{q}{m} - 1 \right]^a}
  \frac{1}{(n\cdot p)^2} &  & 
\end{eqnarray*}
Equation (\ref{2p}) contains all double pole integrals. We have checked that
the result coincides with Appendix 6 of {\cite{soldati}} and  D of {\cite{Lei}}.

Using the same approach we proceed to find the value of the integral for $b$
arbitrary.

 Inspired by the lcg result, equation (\ref{lcg}), we write the ansatz:
\begin{eqnarray}
  \int d p \frac{1}{[p^2 + 2 p\cdot q - m^2]^a} \frac{1}{(n\cdot p)^b} = T (a, b) + &  &\nonumber
  \\
  (- 1)^{a + b} i (\pi)^{\omega} (- 2)^b \frac{\Gamma (a + b - \omega)}{\Gamma
  (a) \Gamma (b)} (F\cdot q)^b \int_0^{^1} d t \frac{t^{b - 1}}{(q\cdot q - 2 F\cdot q n\cdot q t
  + z (F\cdot q)^2 t^2 + m^2)^{a + b - \omega}} \label{T}
\end{eqnarray}
where $T (a, b)$ is a function of $v, w, s, x, y, z$.

Using the identity (\ref{dnmu}) in equation (\ref{T}), we get:
\begin{eqnarray}
	- b \int d p \frac{1}{[p^2 + 2 p\cdot q - m^2]^a} \frac{p_{\mu}}{(n\cdot p)^{b + 1}} =
	\frac{\partial}{\partial n^{\mu}}_{} T (a, b) + &  &  \nonumber\\
	(- 1)^{a + 1} i (\pi)^{\omega} 2^b \frac{\Gamma (a + b - \omega)}{\Gamma (a)
		\Gamma (b)} (F\cdot q)^{b - 1} b E_{\mu \alpha} q^{\alpha} &  &  \nonumber\\
	\int_0^{^1} d t \frac{t^{b - 1}}{(q\cdot q - 2 F\cdot q n\cdot q t + z (F\cdot q)^2 t^2 +
		m^2)^{a + b - \omega}} + &  &  \nonumber\\
	(- 1)^{a + 1} i (\pi)^{\omega} 2^{b + 1} \frac{\Gamma (a + b + 1 -
		\omega)}{\Gamma (a) \Gamma (b)} &  &  \nonumber\\
	(F\cdot q)^b \int_0^{^1} d t \frac{t^{b - 1} (E_{\mu \alpha} q^{\alpha} n\cdot q t -
		F\cdot q q_{\mu} t + n_{\mu} (F\cdot q)^2 t^2 - z t^2 F\cdot q E_{\mu \alpha}
		q^{\alpha})}{(q\cdot q - 2 F\cdot q n\cdot q t + z (F\cdot q)^2 t^2 + m^2)^{a + b + 1 -
			\omega}} &  &  \label{bp1}
\end{eqnarray}
Integrating by parts the coefficient of $E_{\mu \alpha} q^{\alpha}$ in the
second integral, we get:
\begin{eqnarray}
	F \cdot q \int_0^{^1} d t \frac{t^{b - 1} 2 (n \cdot q t - z t^2 F \cdot q)
		(a + b - \omega) (-)}{(q \cdot q - 2 F \cdot q n \cdot q t + z (F \cdot q)^2
		t^2 + m^2)^{a + b + 1 - \omega}} = &  &  \nonumber\\
	- \int_0^{^1} d t t^b \frac{d}{d t} \frac{1}{(q \cdot q - 2 F \cdot q n
		\cdot q t + z (F \cdot q)^2 t^2 + m^2)^{a + b - \omega}} = &  &  \nonumber\\
	- \frac{1}{(q \cdot q - 2 F \cdot q n \cdot q + z (F \cdot q)^2 + m^2)^{a +
			b - \omega}} + &  &  \nonumber\\
	b \int_0^{^1} d t t^{b - 1} \frac{1}{(q \cdot q - 2 F \cdot q n \cdot q t +
		z (F \cdot q)^2 t^2 + m^2)^{a + b - \omega}} &  & \label{52}
\end{eqnarray}
We see that the first integral in equation (\ref{bp1}) is canceled by the second term of
equation (\ref{52}). Finally, we obtain:

\begin{eqnarray}
	- b \int d p \frac{1}{[p^2 + 2 p.q - m^2]^a} \frac{p_{\mu}}{(n.p)^{b + 1}} =
	\frac{\partial}{\partial n^{\mu}}_{} T (a, b)+ (- 1)^{a + b + 1} i (\pi)^{\omega} (- 2)^b
	\frac{\Gamma (a + b - \omega)}{\Gamma (a) \Gamma (b)} &  &  \nonumber\\
	\frac{(F \cdot q)^{b - 1} \left( \frac{x}{D} (F \cdot q) F_{\mu} + \rho
		(q_{\mu} - n_{\mu} F \cdot q - n \cdot q_{} F_{\mu} \right)}{(q \cdot q - 2
		F \cdot q n \cdot q + z (F \cdot q)^2 + m^2)^{a + b - \omega}} + (- 1)^{a +
		1} i (\pi)^{\omega} 2^{b + 1} \frac{\Gamma (a + b + 1 - \omega)}{\Gamma (a)
		\Gamma (b)} &  &  \nonumber\\
	(F \cdot q)^{b + 1} \int_0^{^1} d t \frac{t^b (q_{\mu} - n_{\mu} F \cdot q
		t)}{(q \cdot q - 2 F \cdot q n \cdot q t + z (F \cdot q)^2 t^2 + m^2)^{a + b
			+ 1 - \omega}} &  & \label{bp}
\end{eqnarray}

We have used equation (\ref{Eq}) in the second term of (\ref{bp}).

We also know that:
\begin{eqnarray}
  \frac{\partial}{\partial n^{\mu}}_{} T (a, b) = T_w (a, b) \left( q_{\mu} \frac{1}{\sqrt{z}} - w \frac{n_{\mu}}{z} \right) +
  T_s (a, b) \left( - \sqrt{z} E_{\mu \alpha} q_{\alpha} + s \frac{n_{\mu}}{z}
  \right)
  + T_x (a, b) \bar{n}_{\mu} + 2 n_{\mu} T_z (a, b) &  & 
\end{eqnarray}
We defined: 
\[ T_s = \frac{\partial T}{\partial s}, T_v = \frac{\partial T}{\partial v},T_w = \frac{\partial T}{\partial w}, T_x = \frac{\partial T}{\partial x},T_z = \frac{\partial T}{\partial z}\]

Write, for a Lorentz scalar $h$:
\begin{eqnarray}
  \int d p \frac{1}{[p^2 + 2 p\cdot q - m^2]^{a - 1}} \frac{1}{(n\cdot p)^{b + 1}} = h
  (v, w, s) &  & \nonumber\\
  \frac{\partial}{\partial q^{\mu}} h (v, w, s) = 2 h_v q_{\mu} + h_w
  \frac{n_{\mu}}{\sqrt{z}} + h_s F_{\mu} \sqrt{z} &  & \label{h}
\end{eqnarray}
We use the notation: 
\[ h_s = \frac{\partial h}{\partial s}, h_v = \frac{\partial h}{\partial v},h_w = \frac{\partial h}{\partial w} \]

Picking up the coefficient of $q_{\mu}$ in equations (\ref{bp}) and  (\ref{h}) we get:
\begin{eqnarray}
  h_v = \rho (- 1)^{a + b + 1} i (\pi)^{\omega} (- 2)^b \frac{\Gamma (a + b -
  \omega)}{\Gamma (a - 1) \Gamma (b + 1)} (F\cdot q)^{b - 1} &  & \nonumber \\
  \frac{1}{(q\cdot q - 2 F\cdot q n\cdot q + n\cdot n (F\cdot q)^2 + m^2)^{a + b - \omega}} + &  & \nonumber\\
  (- 1)^{a + b + 1} i (\pi)^{\omega} (- 2)^{b + 1} \frac{\Gamma (a + b + 1 -
  \omega)}{\Gamma (a - 1) \Gamma (b + 1)} &  & \nonumber\\
  (F\cdot q)^{b + 1} \int_0^{^1} d t \frac{t^b}{(q\cdot q - 2 F\cdot q n\cdot q t + n\cdot n (F\cdot q)^2
  t^2 + m^2)^{a + b + 1 - \omega}} + &  & \nonumber\\
  \frac{a - 1}{b} \left( - \rho \sqrt{z} T_s (a, b) + T_w (a, b)
  \frac{1}{\sqrt{z}} \right) &  & 
\end{eqnarray}
Integrating $h_v$ over $v = q\cdot q$,we get:
\begin{eqnarray}
  h = \bar{h} (w, s)+ \rho (- 1)^{a + b} i (\pi)^{\omega} (- 2)^b \frac{\Gamma (a + b - 1 -
  \omega)}{\Gamma (a - 1) \Gamma (b + 1)} (F\cdot q)^{b - 1} &  & \nonumber \\
  \frac{1}{(q\cdot q - 2 F\cdot q n\cdot q + n\cdot n (F\cdot q)^2 + m^2)^{a + b - 1 - \omega}} + &  &\nonumber
  \\
  (- 1)^{a + b} i (\pi)^{\omega} (- 2)^{b + 1} \frac{\Gamma (a + b -
  \omega)}{\Gamma (a - 1) \Gamma (b + 1)} &  &\nonumber \\
  (F\cdot q)^{b + 1} \int_0^{^1} d t \frac{t^b}{(q\cdot q - 2 F\cdot q n\cdot q t + n\cdot n (F\cdot q)^2
  t^2 + m^2)^{a + b - \omega}} + &  & \nonumber\\
  \frac{a - 1}{b} \int d v \left( - \rho \sqrt{z} T_s (a, b) + T_w (a, b)
  \frac{1}{\sqrt{z}} \right)  &  & 
\end{eqnarray}
From this expression we can derive $h_w, h_s$ and compare it with equation (\ref{bp}).
Since the integral exists, the integrability conditions for $h$ are 
satisfied, as was the case for $b = 2$.

To fix $\bar{h} (w, s)$, we impose that the integral  vanishes at $v \to -\infty$(Please see footnote 4). We
get $\bar{h} (w, s)$=0.

Reintroducing the scale invariant variables $v,w,s$ and using equation (\ref{T}), we can write:

\begin{eqnarray}
  T (a - 1, b + 1) = \frac{\rho (- 1)^{a + b} i (\pi)^{\omega} (- 2)^b
  \frac{\Gamma (a + b - 1 - \omega)}{\Gamma (a - 1) \Gamma (b + 1)}
  \frac{1}{z^{\frac{b - 1}{2}}} s^{b - 1}}{(v - 2 s w + s^2 + m^2)^{a + b - 1
  - \omega}} + &  & \nonumber\\
  \frac{a - 1}{b} \int_{- \infty}^v d v' \left( - \rho \sqrt{z} T_s (a, b) + T_w (a, b)
  \frac{1}{\sqrt{z}} \right) &  & \label{T0}
\end{eqnarray}

Notice that the term proportional to $(F\cdot q)^b$ in equation (\ref{T}) is absent from  (\ref{T0}). This says that the ansatz was right.

To simplify the recursion relation we introduce the function $S(a,b)$ as follows:
\begin{equation}
	T (a, b) = \rho (- 1)^{a + b} i (\pi)^{\omega} (- 2)^{b - 1}
	\frac{1}{\Gamma (a) \Gamma (b)} \frac{1}{z^{\frac{b - 2}{2}}} S (a, b) \label{S}
\end{equation}
We get
\begin{eqnarray}
  S (a - 1, b + 1) = \frac{s^{b - 1} \Gamma (a + b - 1 - \omega)}{(v - 2 s w +
  s^2 + m^2)^{a + b - 1 - \omega}} - \frac{1}{2} \int_{- \infty}^v d v' (- \rho z S_s (a, b)
  + S_w (a, b)) &  &  \label{rr}
\end{eqnarray}
subject to the condition $S (a - 1, 1) = 0$.

This recurrence relation is easy to solve because the  $v$ integral is
trivial. Below we write the first terms of the recurrence relation.
\begin{eqnarray}
  S (a - 1, 2) = \frac{\Gamma (a - \omega)}{(v - 2 s w + s^2 + m^2)^{a -
  \omega}} &  & \\
  S (a - 1, 3) = \frac{(s (\rho z + 2) - w \rho z) \Gamma (a + 1 - \omega)}{(v
  - 2 s w + s^2 + m^2)^{a + 1 - \omega}} &  & \\
  S (a - 1, 4) = \frac{(s^2 (3 + 3 \rho z + \rho^2 z^2) - w s (3 \rho z + 2
  \rho^2 z^2) + w^2 \rho^2 z^2) \Gamma (a + 2 - \omega)}{(v - 2 s w + s^2 +
  m^2)^{a + 2 - \omega}} - &  & \\
  \frac{1}{2} \frac{(3 \rho z + \rho^2 z^2) \Gamma (a + 1 - \omega)}{(v - 2 s
  w + s^2 + m^2)^{a + 1 - \omega}} &  & \label{s4} \nonumber
\end{eqnarray}

It is straightforward to write a routine to solve the recurrence relation. We have solved it
using the computer program FORM {\cite{form}}. The  code is written in Appendix C.
 	
As a further check of equation (\ref{2p}) we have calculated the following
finite integral that is not listed neither in {\cite{soldati}} nor {\cite{Lei}}. 

Using equation (\ref{ML}) we get the two-dimensional integral:
\begin{eqnarray}
  \int \frac{d p}{(2 \pi)^2} \frac{1}{[p^2 - m^2]} \frac{1}{(n\cdot p)^2} =
  \frac{i}{4 \pi m^2} \frac{n_0 - n_1}{n_0 n_1 (n_0 + n_1)}=2 \pi \rho \frac{i}{m^2} &  & \label{66}
\end{eqnarray}
We have chosen $n_{\mu} = (n_0, n_1), \bar{n}_{\mu} = (n_0, - n_1)$. We see
that the result coincides with equation (\ref{2p}). Integral (\ref{66}) is calculated in Appendix D.

 In the next section we will find the solution of  recurrence relation equation (\ref{rr}).

\section{Solving the recurrence relation for $S(a,b)$}

Define the function $G(t,b)$ by the following expression:
\begin{equation}
	S (a, b) = \int_0^{\infty} d t G (t, b) t^{a + b - 2 - \omega} e^{- t (v - 2
	s w + s^2 + m^2)}  \label{g}
\end{equation}
Introducing it into equation (\ref{rr}), we see that $G(t,b)$
satisfies the recurrence relation:
\begin{equation}
	G (t, b + 1) = s^{b - 1} + \lambda G (t, b) - \frac{1}{t} \hat{O} G (t,b) \label{g2}
\end{equation}
We have defined:
\begin{eqnarray}
	\hat{O} = \frac{1}{2} (\rho z \partial_s - \partial_w) &  &  \nonumber\\
	\lambda = s (1 + \rho z) - \rho z w &  &  \label{O}
\end{eqnarray}

$\partial_{s(w)}$ denotes the partial derivative respect to  $s(w)$.

Notice that $G(t,b)$ does not depend on $a$.

Equation (\ref{g2}) has  the solution:
\begin{equation}
	G (t,b) = \frac{s^{b - 1} - \lambda^{b - 1}}{s - \lambda} - \sum_{i = 2}^{b -
		2} \left( \lambda - \frac{1}{t} \hat{O} \right)^{i - 2} \frac{\hat{O}}{t}
	\frac{s^{b - i} - \lambda^{b - i}}{s - \lambda},b\ge 1 \label{solg}
\end{equation}

Equation (\ref{solg}) can be proved using induction on $b$.

Notice that $\lambda$ depends on $s,w$, so $\hat{O}$ does not commute with $\lambda$.

Using equation (\ref{solg}), we get:
\begin{eqnarray}
	S (a, 2) = \frac{\Gamma (a + 1 - \omega)}{(v - 2 s w + s^2 + m^2)^{a + 1 -
			\omega}} &  & \\
	S (a, 3) = (s + \lambda) \frac{\Gamma (a + 2 - \omega)}{(v - 2 s w + s^2 +
		m^2)^{a + 2 - \omega}} &  & \\
	S (a, 4) = \frac{s^3 - \lambda^3}{s - \lambda} \frac{\Gamma (a + 3 -
		\omega)}{(v - 2 s w + s^2 + m^2)^{a + 3 - \omega}} - \hat{O} (s + \lambda)
	\frac{\Gamma (a + 2 - \omega)}{(v - 2 s w + s^2 + m^2)^{a + 2 - \omega}} & 
	& \\
	S (a, 5) = \frac{s^4 - \lambda^4}{s - \lambda} \frac{\Gamma (a + 4 -
		\omega)}{(v - 2 s w + s^2 + m^2)^{a + 2 - \omega}} - &  &  \nonumber\\
	\left( \hat{O} \frac{s^3 - \lambda^3}{s - \lambda} + \lambda \hat{O}
	\frac{s^2 - \lambda^2}{s - \lambda} \right) \frac{\Gamma (a + 3 -
		\omega)}{(v - 2 s w + s^2 + m^2)^{a + 1 - \omega}} &  & \\
	S (a, 6) = \frac{s^5 - \lambda^5}{s - \lambda} \frac{\Gamma (a + 5 -
		\omega)}{(v - 2 s w + s^2 + m^2)^{a + 3 - \omega}} - &  &  \nonumber\\
	\frac{\Gamma (a + 4 - \omega)}{(v - 2 s w + s^2 + m^2)^{a + 2 - \omega}}
	\left( \hat{O} \frac{s^4 - \lambda^4}{s - \lambda} + \lambda \hat{O}
	\frac{s^3 - \lambda^3}{s - \lambda} + \lambda^2 \hat{O} (s + \lambda)
	\right) + &  &  \nonumber\\
	\frac{\Gamma (a + 3 - \omega)}{(v - 2 s w + s^2 + m^2)^{a + 1 - \omega}}
	\left( \hat{O}^2 \frac{s^3 - \lambda^3}{s - \lambda} + \hat{O} \lambda
	\hat{O} (s + \lambda) \right) &  & 
\end{eqnarray}
We have verified that equation (\ref{solg}) give the same answers provided by a direct solution of equation (\ref{rr}).

We finally write the main result of this work:
\begin{eqnarray}
	\int dp \frac{1}{[p^2 + 2 p \cdot q - m^2]^a}  \frac{1}{(n \cdot p)^b} =
	\rho (- 1)^{a + b} i (\pi)^{\omega} (- 2)^{b - 1} \frac{1}{\Gamma (a) \Gamma
		(b)} \frac{1}{z^{\frac{b - 2}{2}}} &  &  \nonumber\\
	\int_0^{\infty} d t t^{a + b - 2 - \omega} e^{- t (v - 2 s w + s^2 + m^2)}
	\left( \frac{s^{b - 1} - \lambda^{b - 1}}{s - \lambda} - \sum_{i = 2}^{b -
		2} \left( \lambda - \frac{1}{t} \hat{O} \right)^{i - 2} \frac{\hat{O}}{t}
	\frac{s^{b - i} - \lambda^{b - i}}{s - \lambda} \right) + &  &  \nonumber\\
	(- 1)^{a + b} i (\pi)^{\omega}  (- 2)^b \frac{\Gamma (a + b -
		\omega)}{\Gamma (a) \Gamma (b)}  (F \cdot q)^b  &  &  \nonumber\\
	\int_0^1 dt \frac{t^{b - 1}}{(q \cdot q - 2 F \cdot qn \cdot qt + z
		(F \cdot q)^2 t^2 + m^2)^{a + b - \omega}} &  &  \label{allb}
\end{eqnarray}

We can see that equation (\ref{allb}) preserves naive power counting and goes to the lcg result when $y\to 0,z\to 0$.

\section{Conclusions}

In this paper we have obtained a closed form for the integrals that appear in
loop calculations of gauge models in algebraic non-covariant gauges, outside the lcg. We employed the same
method we used to derive the light-cone integrals in {\cite{alfaro}}.

The procedure is very simple, being based on a scale symmetry plus regularity conditions. For the single spurious pole, the integral satisfies a partial differential equation whose unique regular solution determines the value of the integral. Integrals containing higher order spurious poles are obtained by derivation respect to $n_\mu$ using  (\ref{dnmu}) and then integrating a simple partial differential equation.

The procedure provides  pole and finite parts of the integrals using DR.

We have verified that the integrals obtained in this way coincide with the
ones computed using the generalized ML prescription, equation (\ref{ML}), which 
requires long calculations, having to fix from the beginning a form of
$n_{\mu}, \bar{n}_{\mu}$.

We have also clarified the role of $F_{\mu}$. It must be a null vector, $F\cdot F=0$, in any space-time dimension. We derived this property  from a consistency condition. Lorentz invariance plus $F\cdot F=0$ determines the form of $F_\mu$ in terms of $n_\mu,\bar{n}_\mu$. We have also exhibited a  symmetry of the single spurious pole integral, which provides an independent way to determine $F_\mu$.

Having obtained a closed form for the loop integrals in algebraic non-covariant gauges will greatly facilitate the calculation of Green functions and the discussion of renormalization in these gauges.

Our results provide a robust framework for regularizing integrals in non-covariant gauges. Future work could envisage applications in Very Special Relativity and other non-local gauge theories.

\section{Acknowledgments}

J.A. wants to thank Prof. L.F. Urrutia, Dr. A. Mart\' in and the members of the Instituto de Ciencias Nucleares, UNAM(Mexico), for their kind
hospitality. He also acknowledges the partial support of the  Instituto de F\'isica at PUC.

\section{Appendix A Integrals evaluated at $q_{\mu} = 0$}

In this appendix we write a list of several  integrals that facilitates the comparison with \cite{Lei,soldati}. 

First, we recall some definitions:
\[ \int d p \frac{1}{[p^2 - m^2]^a} \frac{1}{(n\cdot p)^b} = \rho (- 1)^{a + 1} i
   (\pi)^{\omega} (2)^{b - 1} \frac{1}{\Gamma (a) \Gamma (b)}
   \frac{1}{z^{\frac{b - 2}{2}}} S (a, b) |_{q_{\mu} = 0} \nobracket \]
\begin{eqnarray}
  \int d p \frac{1}{[p^2 - m^2]^a} \frac{p_{\mu}}{(n\cdot p)^b} = \rho (- 1)^{a +
  1} i (\pi)^{\omega} 2^{b - 2} \frac{1}{\Gamma (a) \Gamma (b)}
  \frac{1}{z^{\frac{b - 2}{2}}} \frac{\partial S}{\partial q^{\mu}}_{} (a - 1,
  b) |_{q_\mu = 0} \nobracket, & b = 2, 3 \ldots & 
\end{eqnarray}
Below we list some integrals for $b = 1, \ldots .8$ For $b\geq 2$ we rely either on the FORM routine presented in Appendix C
or in Eq.(\ref{allb}).

\begin{equation}
	\int d p \frac{1}{[p^2 - m^2]^a} \frac{p_{\mu}}{(n\cdot p)} = (- 1)^a i
	\pi^{\omega} \frac{\Gamma \left( a - \omega \right)}{\Gamma (a)} 
	\frac{1}{(m^2)^{a - \omega}} F_{\mu}\label{65}
\end{equation}
\begin{eqnarray}
	\int d p \frac{p_{\mu} p_{\nu} p_{\lambda}}{[p^2 - m^2]^a} \frac{1}{n\cdot p} =
	(- 1)^{a + 1} \frac{i \pi^{\omega}}{2} \frac{\Gamma (a - 1 - \omega)}{\Gamma
		(a) (m^2)^{a - 1 - \omega}} &  &  \nonumber\\
	(F_{\mu} F_{\nu} F_{\lambda} z - F_{\mu} F_{\nu} n_{\lambda} - F_{\mu}
	F_{\lambda} n_{\nu} - F_{\lambda} F_{\nu} n_{\mu} + F_{\mu} \eta_{\nu
		\lambda} + F_{\nu} \eta_{\mu \lambda} + F_{\lambda} \eta_{\mu \nu}) &  & 
	\label{3p1}
\end{eqnarray}
\begin{equation}
	\int d p \frac{1}{[p^2 - m^2]^a} \frac{1}{(n\cdot p)^2} = 2 \rho (- 1)^{a + 1} i
	(\pi)^{\omega} \frac{\Gamma (a + 1 - \omega)}{\Gamma (a) (m^2)^{a + 1 -
			\omega}}\label{67}
\end{equation}

\begin{equation}
	\int d p \frac{1}{[p^2 - m^2]^a} \frac{p_{\mu}}{(n\cdot p)^3} = \rho (- 1)^{a +
		1} i (\pi)^{\omega} \frac{\Gamma (a + 1 - \omega)}{\Gamma (a)}
	\frac{(F_{\mu} (\rho z + 2) - n_{\mu} \rho)}{(m^2)^{a + 1 - \omega}}\label{68}
\end{equation}
\begin{equation}
	\int d p \frac{1}{[p^2 - m^2]^a} \frac{1}{(n\cdot p)^4} = \frac{2}{3} \rho^2 (-
	1)^a i (\pi)^{\omega} \frac{\Gamma (a + 2 - \omega)}{\Gamma (a) (m^2)^{a + 2
			- \omega}} (3 + \rho z)\label{69}
\end{equation}

\begin{eqnarray}
	\int d p \frac{1}{[p^2 - m^2]^a} \frac{p_{\mu}}{(n\cdot p)^5} = \rho^2 (- 1)^a i
	(\pi)^{\omega} \frac{\Gamma (a + 2 - \omega)}{6 \Gamma (a) (m^2)^{a + 2 -
			\omega}} &  & \\
	( F_{\mu} (12 + 12\rho z+ 3 \rho^2 z^2)  - n_{\mu} \rho (8 + 3
	\rho z)) &  &  \nonumber
\end{eqnarray}
\begin{eqnarray}
	\int d p \frac{1}{[p^2 - m^2]^a} \frac{1}{(n\cdot p)^6} = \rho^3 (- 1)^{a + 1} i
	(\pi)^{\omega} \frac{1}{15 \Gamma (a)} \frac{\Gamma (a + 3 -
		\omega)}{(m^2)^{a + 3 - \omega}} (20 + 15 \rho z + 3 \rho^2 z^2) &  & 
\end{eqnarray}
\begin{eqnarray}
	\int d p \frac{1}{[p^2 - m^2]^a} \frac{p_{\mu}}{(n\cdot p)^7} = \rho^3 (- 1)^{a +
		1} i (\pi)^{\omega} \frac{1}{90 \Gamma (a)} \frac{\Gamma (a + 3 -
		\omega)}{(m^2)^{a + 3 - \omega}} &  &  \nonumber\\
	(F_{\mu} (120 + 180 \rho z + 90 \rho^2 z^2 + 15 \rho^3 z^3) - n_{\mu} \rho
	(90 + 72 \rho z + 15 \rho^2 z^2)) &  & 
\end{eqnarray}
\begin{eqnarray}
	\int d p \frac{1}{[p^2 - m^2]^a} \frac{1}{(n\cdot p)^8} = \rho^4 (- 1)^a i
	(\pi)^{\omega} \frac{2^7}{\Gamma (a) 7!} \frac{\Gamma (a + 4 -
		\omega)}{(m^2)^{a + 4 - \omega}} &  &  \nonumber\\
	\left( \frac{105}{4} + \frac{63}{2} \rho z + \frac{105}{8} \rho^2 z^2 +
	\frac{15}{8} \rho^3 z^3 \right) &  & 
\end{eqnarray}

We want to mention that there are relations among these integrals, that serve as additional checks of the results of this paper.

Let us recall that:
\begin{eqnarray}
	\rho = \frac{y}{D (x + D)}, & \frac{\partial}{\partial n^{\mu}} \rho =
	\rho^2 n_{\mu} - F_{\mu} \rho (\rho z + 2) &  \label{drho}
\end{eqnarray}
Then
\begin{eqnarray*}
	- \frac{1}{2} \frac{\partial}{\partial n^{\mu}} \int d p \frac{1}{[p^2 -
		m^2]^a} \frac{1}{(n\cdot p)^2} = \int d p \frac{1}{[p^2 - m^2]^a}
	\frac{p_{\mu}}{(n\cdot p)^3} &  & 
\end{eqnarray*}
Using equation (\ref{drho}) we can easily get equation (\ref{68}) from equation (\ref{67}). 

Let us take a second derivative of the equation (\ref{67}), to get:
\begin{equation}
	\int d p \frac{1}{[p^2 - m^2]^a} \frac{p^2}{(n\cdot p)^4} = \frac{1}{3} (- 1)^{a
		+ 1} i (\pi)^{\omega} \frac{\Gamma (a + 1 - \omega)}{\Gamma (a) (m^2)^{a +
			1 - \omega}} \frac{\partial}{\partial n_{\mu}} \frac{\partial}{\partial
		n^{\mu}} \rho 
\end{equation}
But, we have that:
\begin{eqnarray}
	\int d p \frac{1}{[p^2 - m^2]^a} \frac{p^2}{(n\cdot p)^4} = \int d p
	\frac{1}{[p^2 - m^2]^{a - 1}} \frac{1}{(n\cdot p)^4} + \frac{m^2}{a - 1}
	\partial_{m^2} \int d p \frac{1}{[p^2 - m^2]^{a - 1}} \frac{1}{(n\cdot p)^4}
\end{eqnarray}
The solution of the differential equation for $\int d p \frac{1}{[p^2 -
	m^2]^{a - 1}} \frac{1}{(n\cdot p)^4}$ that preserves the scaling in $m^2$ is:
\begin{eqnarray}
	\int d p \frac{1}{[p^2 - m^2]^{a - 1}} \frac{1}{(n\cdot p)^4} = \frac{1}{3} (-
	1)^{a + 1} i (\pi)^{\omega} \frac{\Gamma (a + 1 - \omega)}{\Gamma (a - 1)
		(\omega - 2)} (m^2)^{\omega - 2 + 1 - a} \frac{\partial}{\partial n_{\mu}}
	\frac{\partial}{\partial n^{\mu}} \rho &  & 
\end{eqnarray}
Using that:
\begin{equation}
	 \frac{\partial}{\partial n_{\mu}} \frac{\partial}{\partial n^{\mu}} \rho =
	(d - 4) \rho^2 (\rho z + 3) 
\end{equation}
we get agreement with equation (\ref{69}).

In this way we can generate all integrals listed above, starting from equation (\ref{65}).

In addition to the previous relations, a symmetric $E_{\mu \nu}$ implies that  $F_\mu$ is the gradient of a potential, $F_{\mu} = \phi_{, \mu}$. Here $\phi_{, \mu}=\frac{\partial\phi}{\partial n^\mu}$. We get:
\begin{equation}
	\phi = \log \left( \frac{x + D}{\sqrt{y}} \right)
\end{equation}

Therefore,we can write:
\begin{eqnarray}
	\int d p \frac{1}{(p^2 - m^2 + i \varepsilon)^a} \frac{p_{\mu}}{n\cdot p} = A_a
	F_{\mu} &  & \\
	A_a =  (- 1)^a i
	\pi^{\omega} \frac{\Gamma \left( a - \omega \right)}{\Gamma (a)} 
	\frac{1}{(m^2)^{a - \omega}} &  & 
\end{eqnarray}
This permits to prove that:
\begin{eqnarray}
	\int d p \frac{1}{(p^2 - m^2 + i \varepsilon)^a} \frac{p_{\mu}
		p_{\nu}}{(n\cdot p)^2} = A_a E_{\mu \nu}, & E_{\mu \nu} = - \phi_{, \mu \nu} & \\
	\int d p \frac{1}{(p^2 - m^2 + i \varepsilon)^a} \frac{p_{\mu} p_{\nu}
		p_{\lambda}}{(n\cdot p)^3} = \frac{1}{2} \phi_{, \mu \nu \lambda} A_a &  & \\
	\int d p \frac{1}{(p^2 - m^2 + i \varepsilon)^a} \frac{p_{\mu_1} \ldots
		p_{\mu_n}}{(n\cdot p)^b} = \frac{(- 1)^{b + 1}}{\Gamma (b)} \phi_{, \mu_1 \ldots
		. \mu_b} A_a &  & 
\end{eqnarray}

\section{Appendix B: A symmetry of single pole integrals}

We want to mention that $F_{\mu}$ is invariant under the following
infinitesimal transformations:
\begin{eqnarray}
  \delta n_{\mu} = 0 &  &  \nonumber\\
  \delta \bar{n}_{\mu} = c \frac{n_{\mu}}{n\cdot n} + d \bar{n}_{\mu} &  & 
  \label{sym}
\end{eqnarray}
Here $c, d$ are arbitrary parameters.

This means that equation (\ref{onep}) is invariant under (\ref{sym}).

This symmetry determines $F_{\mu}$ uniquely, without imposing $F\cdot F = 0$.

Higher poles integrals do not respect the symmetry corresponding to the
infinitesimal parameter $c$. In fact $\delta \rho = \frac{c}{z D}$. $D$ is
defined in equation (\ref{Fmu}).

From equation (\ref{ML}) it is easy to see that the scaling defined by the
parameter $d$ is always a symmetry.

\section{Appendix C: solving equation (\ref{rr}) using FORM}

\begin{verbatim}
	***********************rec.frm*******************************
	
	#-;
	
	F Ts,Tw,s,w,int,q;
	
	S rho,a,x,b,z,om;
	
	*q includes the gamma factor in the numerator
	
	Off statistics;
	
	.global
	
	
	GL t2=q(a-om);
	
	p;
	
	.store
	
	* Change next line to get other b's
	
	#do i=2,10
	
	GL t'i'2=(-rho*z*Ts+Tw)*t'i';
	
	id q(a?)=-int*q(a+1)/2;
	
	repeat;
	
	id int*q(a?)=-q(a-1);
	
	id Ts*q(a?)=(2*w(1)-2*s(1))*q(a+1);
	
	id Tw*q(a?)=(2*s(1))*q(a+1);
	
	id Ts*s(a?)=a*s(a-1)+s(a)*Ts;
	
	id Ts*w(a?)=w(a)*Ts;
	
	id Tw*s(a?)=s(a)*Tw;
	
	id Tw*w(a?)=a*w(a-1)+w(a)*Tw;
	
	id s(a?)*w(b?)=w(b)*s(a);
	
	id s(a?)*s(b?)=s(a+b);
	
	id w(a?)*w(b?)=w(a+b);
	
	id w(0)=1;
	
	id s(0)=1;
	
	endrepeat;
	
	.store
	
	GL t{'i'+1}=s('i'-1)*q(a+'i'-1-om)+t'i'2;
	
	p;
	
	.store
	
	#enddo
	
	.end
\end{verbatim}

We have defined:
\begin{eqnarray*}
	q (a - \tmop{om}) = \frac{\Gamma (a - \omega)}{(v - 2 s w + s^2 + m^2)^{a -
			\omega}}, & t' i' = S (a - 1, i) & \\
	s (n) = s^n, & w (n) = w^n, & \tmop{rho} = \rho
\end{eqnarray*}
\section{ Appendix D: Some finite integrals}
In this appendix, we calculate the integrals defined in equations (\ref{38},\ref{39},\ref{66}) using equation (\ref{ML}).

To calculate the integrals, we see that the even functions of $p_0$ and $p_1$
survive. Then we introduce Feynman parameter $x$ to write:
\begin{eqnarray}
	\int \frac{d p}{(2 \pi)^2} \frac{1}{[p_0^2 - p_1^2 - m^2 + i \varepsilon]^2}
	p_0 \frac{n_0 p_0 + n_1 p_1}{(n_0 p_0)^2 - (n_1 p_1)^2 + i \varepsilon} = & 
	&  \nonumber\\
	2 n_0 \int_0^1 d x x \int \frac{d p}{(2 \pi)^2} \frac{p_0^2}{[p_0^2 x -
		p_1^2 x - m^2 x + (1 - x) (n_0^2 p_0^2 - n_1^2 p_1^2) + i \varepsilon]^3} =
	&  &  \nonumber\\
	2 n_0 \int d x \frac{x}{(x + (1 - x) n_0^2)^3} \int \frac{d p}{(2 \pi)^2}
	\frac{p_0^2}{\left[ p_0^2 - p_1^2 \frac{x}{A (x)} - m^2 \frac{x}{A (x)} -
		\frac{(1 - x)}{A (x)} n_1^2 p_1^2 + i \varepsilon \right]^3} = &  & 
	\nonumber\\
	n_0 \frac{i}{\sqrt{4 \pi}} \int_0^1 d x \frac{x}{(x + (1 - x) n_0^2)^3} \int
	\frac{d p_1}{(2 \pi)} \frac{\Gamma (3 / 2)}{2} \frac{1}{\left( p_1^2
		\frac{x}{A (x)} + m^2 \frac{x}{A (x)} + \frac{(1 - x)}{A (x)} n_1^2 p_1^2 -
		i \varepsilon \right)^{3 / 2}} = &  &  \nonumber\\
	n_0 \frac{i}{2^3} \int_0^1 d x \frac{x}{(x + (1 - x) n_0^2)^3} \frac{1}{B
		(x)^{3 / 2}} \int \frac{d p_1}{(2 \pi)} \frac{1}{\left( p_1^2 + m^2
		\frac{x}{B (x) A (x)} \right)^{3 / 2}} = &  &  \nonumber\\
	n_0 \frac{i}{2^3} \int d x \frac{x}{(x + (1 - x) n_0^2)^3} \frac{1}{B (x)^{3
			/ 2}} \left( m \sqrt{\frac{x}{B (x) A (x)}} \right)^{- 2} \frac{1}{\pi} = & 
	&  \nonumber\\
	\frac{i n_0}{8 \pi m^2} \int_0^1 d x \frac{1}{\sqrt{(x + (1 - x) n_0^2)^3}}
	\frac{1}{\sqrt{x + (1 - x) n_1^2}} = &  &  \nonumber\\
	\frac{i n_0}{8 \pi m^2} \left( - \frac{2}{n_0  (n_0^2 - n_1^2)} (n_1 - n_0)
	\right) = \frac{i n_0}{4 \pi m^2} \frac{1}{n_0 (n_0 + n_1)} &  & 
\end{eqnarray}
We have defined:
\begin{eqnarray}
	A (x) = x + (1 - x) n_0^2 &  & \\
	B = \frac{x}{A (x)} + \frac{(1 - x)}{A (x)} n_1^2 &  & 
\end{eqnarray}
That is, we get:
\begin{equation}
	\int d p \frac{1}{[p^2 - m^2]^2} \frac{p_0}{n.p} = \frac{i n_0 \pi}{m^2}
	\frac{1}{n_0 (n_0 + n_1)}
\end{equation}
Similarly, we get:
\begin{eqnarray}
	\int \frac{d p}{(2 \pi)^2} \frac{1}{[p^2 - m^2]^2} \frac{p_1}{n.p} = &  & 
	\nonumber\\
	\int \frac{d p}{(2 \pi)^2} \frac{1}{[p_0^2 - p_1^2 - m^2 + i \varepsilon]^2}
	p_1 \frac{n_0 p_0 + n_1 p_1}{(n_0 p_0)^2 - (n_1 p_1)^2 + i \varepsilon} = & 
	&  \nonumber\\
	\int \frac{d p}{(2 \pi)^2} \frac{1}{[p_0^2 - p_1^2 - m^2 + i \varepsilon]^2}
	\frac{n_1 p_1^2}{(n_0 p_0)^2 - (n_1 p_1)^2 + i \varepsilon} = &  & 
	\nonumber\\
	2 n_1 \int_0^1 \frac{d x}{A^3} x \int \frac{d p}{(2 \pi)^2}
	\frac{p_1^2}{\left[ p_0^2 - p_1^2 \frac{x}{A (x)} - m^2 \frac{x}{A (x)} -
		\frac{(1 - x)}{A (x)} n_1^2 p_1^2 + i \varepsilon \right]^3} = &  & 
	\nonumber\\
	2 n_1 \frac{1}{(2 \pi)} \frac{- i}{\sqrt{4 \pi}} \frac{\Gamma (5 / 2)}{2}
	\int_0^1 \frac{d x}{A^3 B^{5 / 2}} x \int d p_1 \frac{p_1^2}{\left( p_1^2 +
		m^2 \frac{x}{A (x) B} \right)^{5 / 2}} = &  &  \nonumber\\
	2 n_1 \frac{1}{(2 \pi)} \frac{- i}{\sqrt{4 \pi}} \frac{\Gamma (5 / 2)}{2}
	\frac{2}{3} \int_0^1 \frac{d x}{A^3 B^{5 / 2}} \frac{x}{m^2 \frac{x}{A (x)
			B}} = &  &  \nonumber\\
	- i \frac{n_1}{8 \pi m^2} \int_0^1 d x \frac{1}{\sqrt{(x + (1 - x)
			n_1^2)^3}} \frac{1}{\sqrt{x + (1 - x) n_0^2}} = &  &  \nonumber\\
	- i \frac{n_1}{8 \pi m^2} \left( - \frac{2}{n_1  (n_0^2 - n_1^2)} (n_1 -
	n_0) \right) = - i \frac{n_1}{4 \pi m^2} \left( \frac{1}{n_1 (n_0 + n_1)}
	\right) &  & 
\end{eqnarray}
Therefore, we get:
\begin{equation}
	\int d^{} p \frac{1}{[p^2 - m^2]^2} \frac{p_1}{n.p} = - i \frac{n_1
		\pi}{m^2} \left( \frac{1}{n_1 (n_0 + n_1)} \right)
\end{equation}

Now we wish to calculate the following integral:
\begin{eqnarray}
	\int \frac{d p}{(2 \pi)^2} \frac{1}{[p^2 - m^2]} \frac{1}{(n.p)^2} = \int
	\frac{d p}{(2 \pi)^2} \frac{1}{[p_0^2 - p_1^2 - m^2 + i \varepsilon]}
	\frac{(n_0 p_0 + n_1 p_1)^2}{(n_0^2 p_0^2 - n_1^2 p_1^2 + i \varepsilon)^2}
	= &  &  \nonumber\\
	2 \int_0^1 d x (1 - x) \int \frac{d p}{(2 \pi)^2} \frac{(n_0 p_0 + n_1
		p_1)^2}{[(p_0^2 - p_1^2 - m^2) x + (1 - x) (n_0^2 p_0^2 - n_1^2 p_1^2) + i
		\varepsilon]^3} = &  &  \nonumber\\
	2 \int_0^1 d x (1 - x) \int \frac{d p}{(2 \pi)^2} \frac{(n_0^2 p_0^2 + n_1^2
		p_1^2)}{[(p_0^2 - p_1^2 - m^2) x + (1 - x) (n_0^2 p_0^2 - n_1^2 p_1^2) + i
		\varepsilon]^3} = &  &  \nonumber\\
	2 \int_0^1 d x (1 - x) \int \frac{d p_1}{(2 \pi)} \frac{n_0^2}{A^3}
	\frac{i}{2^4} \frac{1}{\left[ (p_1^2 + m^2) \frac{x}{A} + n_1^2 p_1^2
		\frac{1 - x}{A} \right]^{3 / 2}} + &  &  \nonumber\\
	2 \int_0^1 d x (1 - x) \int \frac{d p_1}{(2 \pi)} \frac{n_1^2}{A^3} \frac{-
		i}{\sqrt{4 \pi}} \frac{\Gamma (5 / 2)}{2} \frac{p_1^2}{\left[ (p_1^2 + m^2)
		\frac{x}{A} + n_1^2 p_1^2 \frac{1 - x}{A} \right]^{5 / 2}} = &  & 
	\nonumber\\
	\frac{i}{2^3 m^2 \pi} \int_0^1 d x \frac{(1 - x)}{x} \left\{
	\frac{n_0^2}{A^2 B^{1 / 2}} - \frac{n_1^2}{A^2 B^{3 / 2}} \right\} = &  & 
	\nonumber\\
	\frac{i}{2^3 m^2 \pi} \int_0^1 d x \frac{x}{1 - x} &  &  \nonumber\\
	\left( \frac{n_0^2}{(x + (1 - x) n_0^2)^{3 / 2} (x + (1 - x) n_1^2)^{1 / 2}}
	- \frac{n_1^2}{(x + (1 - x) n_0^2)^{1 / 2} (x + (1 - x) n_1^2)^{3 / 2}}
	\right) = &  &  \nonumber\\
	\frac{i}{2^3 m^2 \pi} 2 \frac{n_0^2 + n_1^2 - 2 n_0 n_1}{n_0 n_1 (n_0^2 -
		n_1^2)} = \frac{i}{4 \pi m^2} \frac{n_0 - n_1}{n_0 n_1 (n_0 + n_1)} &  & 
\end{eqnarray}
Therefore, we get:
\begin{equation}
	\int d p \frac{1}{[p^2 - m^2]} \frac{1}{(n.p)^2} = \frac{i \pi}{m^2}
	\frac{n_0 - n_1}{n_0 n_1 (n_0 + n_1)}
\end{equation}

On the other hand, we have that:
\begin{eqnarray}
	x = n_0^2 + n_1^2, & y = z = n_0^2 - n_1^2 &  \nonumber\\
	D = 2 n_0 n_1, & x + D = (n_0 + n_1)^2 &  \nonumber\\
	\rho = \frac{n_0 - n_1}{2 n_0 n_1 (n_0 + n_1)} &  & 
\end{eqnarray}
and
\begin{eqnarray}
	F_0 = \frac{1}{n_0 + n_1} &  &  \nonumber\\
	F_1 = - \frac{1}{n_0 + n_1} &  & 
\end{eqnarray}
We can verify that equations (\ref{38},\ref{39},\ref{66}) are correct.
\section{Appendix E: Comparison with \cite{Lei,soldati}}

In this section, we want to compute some of the integrals listed in  \cite{Lei,soldati}, to  provide a test of our results.

\begin{eqnarray}
	\mathbb{P} \int \frac{d p}{(n \cdot p)^2} \frac{1}{((p - q)^2 - m^2)} = 2
	\rho \frac{i \pi^2}{2 - \omega} &  & 
\end{eqnarray}
This is A6.36 of \cite{soldati}. It is obtained from equation (\ref{2p}) evaluated at $a=1$.

To compute the pole part in the next integral, we expand:
\begin{eqnarray}
	\frac{1}{(p - q)^2 - m^2} = \frac{1}{p^2 - m^2} \left( 1 + 2 \frac{p \cdot
		q}{p^2 - m^2} + o (q^2) \right) &  & \label{100}
\end{eqnarray}

\begin{eqnarray}
	\mathbb{P} \int \frac{d p}{n \cdot p} \frac{p_{\mu} p_{\nu}}{((p - q)^2 -
		m^2) p^2} = 2 q^{\lambda} \mathbb{P} \int \frac{d p}{n \cdot p}
	\frac{p_{\mu} p_{\nu} p_{\lambda}}{(p^2 - m^2)^3} = \frac{i \pi^2}{2 -
		\omega} \frac{1}{2} &  & \nonumber\\
	\{ F_{\mu} F_{\nu} F \cdot q z - F_{\mu} F_{\nu} n \cdot q - F \cdot q
	(F_{\mu} n_{\nu} + F_{\nu} n_{\mu}) + F_{\mu} q_{\nu} + F_{\nu} q_{\mu} +
	\nobracket F \cdot q \eta_{\mu \nu} \} \nobracket &  & \label{111}
\end{eqnarray}

The contribution to the integral of the first term in the expansion (\ref{100}) vanishes because the integrand is odd in $p_\mu$.
To obtain integral (\ref{111}) we used equation (\ref{2p3}). We agree with Appendix D of \cite{Lei}. Notice that $F_\mu=\frac{F_\mu^L}{n\cdot F^L}$, where $F_\mu^L$ is  Leibbrandt's $F_\mu$.

From equation (\ref{2p}) we get:
\begin{eqnarray}
	\int d p \frac{1}{[p^2 - m^2]^a} \frac{(p\cdot q)^4}{(n\cdot p)^2} = &  & \nonumber\\
	\frac{3}{2} (- 1)^{a + 1} i \pi^{\omega} \frac{\Gamma (a - 1 -
		\omega)}{\Gamma (a) (m^2)^{a - 1 - w}} \left( \rho U^2 + (F\cdot q)^2 \left( 2
	q\cdot q - \frac{8}{3} F\cdot q n\cdot q + z (F\cdot q)^2 \right) \right) &  & \nonumber\\
	U = q \cdot q - 2 F \cdot q n \cdot q + z (F \cdot q)^2 \ \  \label{sol4}
\end{eqnarray}
For $a=3$ this equation provides the pole part of the integral written in   A6.45 of \cite{soldati}. Additionally, it gives the value of  the whole integral for  arbitrary $a$.

Finally, we compute integral (\ref{1}) using ML:
\begin{eqnarray}
	I = \int \frac{d p}{p^2 (p - k)^2 n\cdot p} = \int_0^1 d x \int \frac{d p}{n\cdot p
		[p^2 - 2 k\cdot p x + x k^2]^2} = &  &\nonumber \\
	\int_0^1 d x 2 i \pi^2 \bar{n} \cdot (- k x) \int_0^1 d t \frac{1}{(- x k^2 +
		k^2 x^2 - 2 n \cdot (- k x) (\bar{n} \cdot (- k x)) t)} = &  & \nonumber\\
	i \pi^2 \frac{1}{n \cdot k} \int_0^1 d x \frac{1}{x} \log \left( 1 + \frac{2
		n \cdot k \bar{n} \cdot k x}{(1 - x) k^2} \right) = &  &\nonumber \\
	i \pi^2 \frac{1}{n \cdot k} \left( \frac{\pi^2}{6} + \int_{_0}^{1 - \frac{2
			n\cdot k \bar{n} \cdot k}{k^2}} d u \frac{\log (1 - u)}{u} \right) &  & 
\end{eqnarray}
We have used equation (\ref{lcg1}) with $q_{\mu} = - k_{\mu} x, m^2 = - x k^2$. The integral is finite and agree with equation (6.5) of \cite{soldati}.
\end{document}